# ANALYSIS OF SECURITY THREATS IN WIRELESS SENSOR NETWORK

Sahabul Alam ˋ and Debashis De

Department of Computer Science and Engineering, West Bengal University of Technology, Kolkata, India

*ABSTRACT*

*Wireless Sensor Network(WSN) is an emerging technology and explored field of researchers worldwide in the past few years, so does the need for effective security mechanisms. The sensing technology combined with processing power and wireless communication makes it lucrative for being exploited in abundance in future. The inclusion of wireless communication technology also incurs various types of security threats due to unattended installation of sensor nodes as sensor networks may interact with sensitive data and /or operate in hostile unattended environments. These security concerns be addressed from the beginning of the system design. The intent of this paper is to investigate the security related issues in wireless sensor networks. In this paper we have explored general security threats in wireless sensor network with extensive study.*

*KEYWORDS*

*Wireless Sensor Networks,Security,Threats,Attacks*

## 1. INTRODUCTION

Wireless sensor networks are network of thousand of sensor nodes. Sensor nodes are small in size, less memory space, cheaper in price with restricted energy source and limited processing capability. WSNs are rapidly gaining popularity due to low cost solutions to a variety of real world challenges [1].The basic idea of sensor network is to disperse tiny sensing devices, which are capable of sensing some changes of incidents / parameters and communicating with other devices over a specific geographic area for some specific purposes like surveillance, environmental monitoring, target tracking etc. Sensor can monitor pressure,humidity, temperature, vehicular movement, lightning conditions, mechanical stress levels on attached objects and other properties [2].Due to the lack of data storage and power sensor networks introduce severe resource constraints. These are the obstacles to the implementation of traditional computer security techniques in a WSN. Security defenses harder in WSN due to the unreliable communication channel and unattended operation. As a result these networks require some unique security policies.Cryptography,steganography and other basics of network security and their applicability can be used to address the critical security issues in WSN. Many researchers have begun to address of maximizing the processing capabilities and energy saving of sensor nodes with securing them against attackers.There are different types of attacks designed to exploit the unreliable communication channels and unattended operation of WSNs. Physical attacks to sensors play an important role in the operation of WSNs due to the inherent unattended feature. We explore various types of attacks and threats against WSN.

## 2. FUNDAMENTAL SECURITY SCHEMES IN WIRELESS SENSOR NETWORK

Security encompasses the characteristics of authentication, privacy,integrity,anti-playback and nonrepudiation. The more the risk of secure transmission of information over the network has increased as the more dependency on the information provided by the networks has been increased.Several cryptographic, steganographic and other techniques are used which are well known for the secure transmission of various types of information over networks [3].In this section, we discuss the network security basic and how the techniques are meant for wireless sensor networks.

### 2.1 Cryptography

The encryption-decryption techniques devised for the traditional wired networks are not feasible to be applied directly for wireless sensor networks. WSNs consist of tiny sensors which really suffer from the lack of battery power,processing and memory. Any encryption scheme applying on WSNs require transmission of extra bits,hence extra processing,memory and battery power which are very important resources for the sensor's longevity. Applying the security mechanisms such as encryption could also increase delay, jitter and packet loss in wireless sensor networks [3]. There are some key questions arise when applying encryption schemes to WSNs like, how the keys are generated or disseminated. There is an important issue how the keys could be modified time to time for encryption as there is minimal (or no) interaction for the sensors. There are other many issues how keys are revoked, assigned to a new sensor added to the network or renewed for ensuring robust security for the network. There could not be an efficient solution for adopting of pre-loaded keys or embedded keys.

### 2.2 Steganography

Steganography aims at hiding the existence of the message while cryptography aims at hiding the content of a message. In steganography a message embeds into the multimedia data(image,sound,video etc.).Steganography modifies the carrier in a way that is not perceptible and looks just like ordinary. It is very useful when we want to distribute secret data publicly and in the case that we want to send a secret data without sender information as it hides the existence of the covert channel. Securing wireless sensor networks is not directly related to steganography and processing multimedia data (like audio,video). With the inadequate resources of the sensors for securing wireless sensor networks is difficult and an open research issue.

### 2.3 Physical Layer Secure Access

Frequency hopping provides physical layer secure access in WSN. A dynamic combination of the parameters like hopping set(available frequencies for hopping), hopping pattern (the sequence in which the frequencies from the available hopping set is used) and dwell time(time interval per hop) could be used with a little expense of memory, processing and energy resources. In physical layer secure access the efficient design is required so that the hopping sequence is modified in less time than to discover it and for employing this both the sender and receiver should maintain a synchronized clock. A scheme as proposed in [4] could also be utilized which introduces secure physical layer employing the singular vectors with the channel synthesized modulation.

## 3. THREAT MODEL

It is usually assumed that an attacker may know the security mechanisms that are deployed in a sensor network. Attackers may be able to compromise a node or even physically capture a node. Most WSN nodes are viewed as non-tamper resistant due to the high cost of deploying tamper-resistant sensor nodes. The attacker is capable of stealing the key materials contained within the compromised node. Base stations are regarded as trustworthy in WSNs. Most researchers focus on secure routing between sensors and the between base stations.

Attacks in sensor networks can be classified into the following categories

1. Outsider Vs. insider attacks: Outsider attacks are attacks from nodes which do not belong to a WSN. Insider attacks occur when legitimate nodes of a WSN behave in unintended or unauthorized ways.

2. Passive Vs. active attacks: Passive attacks include eavesdropping on or monitoring packets exchanged with in a WSN. Active attacks involve some modifications of the data stream or the creation of a false stream.

3. Mote-class Vs. Laptop-class attacks: An adversary attacks a WSN by using a few nodes with similar capabilities to the network nodes in mote-class attacks. Mote class attackers can jam the radio link in it's immediate vicinity. In laptop class attacks an adversary can use more powerful devices (e.g a laptop) to attack a WSN. These devices have greater transmission range, energy reserves and processing power than the network nodes. A laptop class attacker might be able to eavesdrop on an entire network.

## 4. THREAT ATTACKS IN WIRELESS SENSOR NETWORK

Why is security necessary in WSN? Due to the broadcast nature of the transmission medium wireless sensor networks are vulnerable. There are another reason of vulnerability of WSNs are nodes are often placed in a hostile or dangerous environment and they are not physically safe. Most of the threats and attacks against security in wireless sensor networks are almost similar to their wired counterparts while some are exacerbated with the inclusion of wireless connectivity. WSNs are usually more vulnerable to various security threats because the unguided transmission medium is more susceptible to security attacks, but also through traffic analysis, privacy violation, physical attacks and so on.

Attacks on WSNs can be classified from two different levels of views

1. Attack against security mechanisms
2. Attack against basic mechanisms (like routing mechanisms)

In many applications the data obtained by the sensing nodes need to be authentic [5]. A false or malicious node could intercept private information in the absence of proper security or could send false messages to nodes in the network. Different possible attacks can be categorized as follows

### 4.1. Denial of Service Attacks

In WSN, Denial of Service (DOS) is produced by the unintentional failure of nodes or malicious action. In DOS attack the adversary attempts to subvert, disrupt or destroy a network. DOS attack diminishes a network capability to provide a service for any event. The simplest DOS attack tries to exhaust the resources available to the victim node, by sending extra unnecessary packets and thus prevents legitimate network users from accessing services or resources to which they are entitled [6].

### 4.1.1 Jamming

Jamming is a DOS attack at physical layer. Jamming interferes with the radio frequencies that a network's nodes are using [7]. A jamming source may either be powerful enough to disrupt the entire network or less powerful and only able to disrupt a smaller portion of the network. It creates radio interference and resource exhaustion.To defend against jamming involve variations of spread-spectrum communication such as frequency hopping and code spreading.Security class of this attack is modification.

### 4.1.2 Tampering

Another DOS attack in physical layer is tampering. By physical access an attacker can extract sensitive information such as cryptographic keys or other data on the node. A compromised node creates, which the attacker controls by altering or replacing node. Vulnerability of this attack is logical.One defence to this attack involves tamper-proofing the node's physical package.

### 4.1.3 Collisions

Collision is a DOS attack in the data link layer. When two nodes attempt to transmit on the same frequency simultaneously a collision occurs. A change will likely to occur in the data portion when packets collide and causing a checksum mismatch at the receiving end. The packet will then be discarded as invalid. An adversary may strategically cause collisions in specific packets such as ACK control messages. Error-correcting codes use to defend against collisions.

### 4.1.4 Exhaustion

It is another type of DOS attack in link layer. An attacker can use repeated collisions to cause resource exhaustion. For example, a native link-layer implementation may continuously attempt to retransmit the corrupted packets. The energy reserves of the transmitting node unless these hopeless retransmissions are discovered or prevented. Applying rate limits to the MAC admission control is a possible solution of exhaustion.

### 4.1.5 Unfairness

Unfairness is a weak of a DOS attack in link layer. An attacker, may cause unfairness in a network by using the above link- layer attacks. Instead of preventing access to a service outright, an attacker can degrade it in order to gain advantage such as causing other nodes in a real time MAC protocol to miss their transmission deadline.

### 4.1.6 Flooding

Flooding is a DOS attack in transport layer. A protocol becomes vulnerable to memory exhaustion through flooding when it maintains at either end of a connection. An attacker may repeatedly make new connection requests until the resources required by each connection are exhausted or reach a maximum limit. In either case, further legitimate requests will be ignored. Disrupt communication is one of purpose of this attack. It creates resource exhaustion and reduces availability.

#### 4.1.7 Desynchronization

Disruption of an existing connection is desynchronization. For example, an attacker may repeatedly spoof messages to an end host, causing that host to request the retransmission of missed frames. With proper timing, an attacker may degrade or even prevent the ability of the end hosts to successfully exchange data. A possible solution to this type of attack is to require authentication of all packets communicated between hosts. The authentication method would be secure as an attacker will be unable to send the spoofed messages to the end hosts.Vulnerability of this attack is logical.

#### 4.1.8 Data Integrity Attack

Data integrity attacks are caused by changing the data contained within the packets or injecting false node while the data travelling among the nodes in WSN. The attacker node must have more processing, memory and energy than the sensor nodes. The goals of this attack are to falsify routing data in order to disrupt the sensor network's normal operation and also falsifies sensor data by doing so compromise the victim's research. Digital signatures and asymmetric key systems are used to defend against this attack. This requires a lot of additional overhead and is difficult to adapt in WSN.

### 4.2 The Sybil Attack

The sensors in a WSN might need to work together to accomplish a task in many cases, hence they can use distribution of subtasks and redundancy of information. In such case a single node duplicates itself and presented in the multiple locations. The Sybil attack targets fault tolerant schemes such as multipath routing, distribute storage and topology maintenance. Sybil attack tries to degrade the integrity of data security and resource utilization that the distributed algorithm attempts to achieve. Any peer to peer network especially wireless adhoc network is vulnerable to sybil attack. Encryption and authentication techniques can prevent an outsider to launch a sybil attack on the sensor network. If a compromised node pretends to be two of the three nodes the algorithms used may conclude that redundancy has been achieved while in reality it has not. Public key cryptography can prevent an insider attack but it is too expensive to be used in the resource constrained sensor network. Figure 1 shows sybil attack.

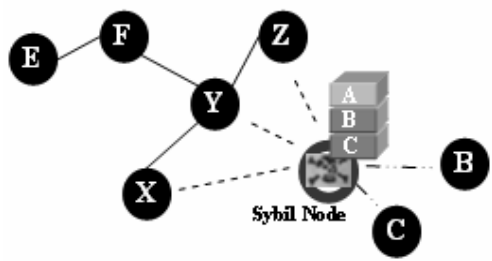

Figure 1. Sybil Attack

### 4.3 Blackhole Attack

A malicious node acts as a blackhole[8] in the range of the sink attracts the entire traffic to be routed through it by advertising itself as the shortest route. The adversary drops packets coming from specific sources in the network. Once the malicious device is in between the communicating nodes (for example, sink and sensor node), it is able to do anything with the packets passing between them. This attack can also affect the nodes those are considerably far from the base stations. It creates high rate of packet loss,network partition.It decreases the throughput of a subset of nodes.The network architecture of this attack is traditional wireless sensor network. Figure 2 shows the conceptual view of a blackhole attack.

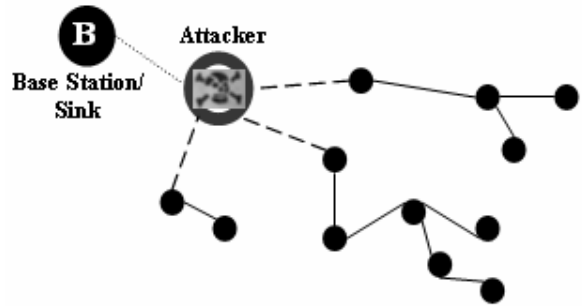

Figure 2. Blackhole Attack

### 4.4 HELLO Flood Attack

HELLO Flood is a novel attack against sensor networks. It is introduced in [9].This attack uses HELLO packets as a weapon to convince the sensors in WSN. Many protocols require nodes to broadcast HELLO packets to announce themselves to their neighbours and a receiving node may assume that it is within normal radio range of the sender. This assumption may be false because a laptop class attacker broadcasts routing or other information with large enough transmission power could convince every node in the network that the adversary is it's neighbour. For example, an adversary advertises a very high quality route to the base station to every node in the network could cause a large number of nodes to attempt to use this route, but those nodes sufficiently far away from the adversary would be sending packets into oblivion. The network is left in a state of confusion.This attack wastes node's energy due to laptop class attacker and creates routing delay. The network architecture of this attack is traditional wireless sensor network.

### 4.5 Traffic Analysis Attack

WSNs consist of many low-power sensors communicating with a few powerful and robust base stations. Data gathered by the individual nodes routed to the base station. Often, for an adversary to effectively render the network useless, the attacker can simply disable the base station. Even when encrypted messages are transferred, it still leaves a high possibility analysis of the communication patterns. Sensor activities can potentially reveal enough information to enable an adversary to cause malicious harm to the sensor network. Table 1shows layer based attacks and possible security mechanisms in wireless sensor networks.

| Layer | Attacks | Security Approach Defenses |
|---|---|---|
| Physical layer | Jamming | Lower duty cycle, Priority messages, Spread-spectrum techniques |
| | Tampering | Tamper proofing, Hiding |
| Data Link layer | Collision | Error-correcting code |
| | Exhaustion | Rate limitation |

| | Unfairness | Small frames |
|---|---|---|
| Network layer | Spoofed, altered or replayed routing information | Authentication, Monitoring |
| | Selective forwarding | Probing, Redundancy |
| | Sink hole | Monitoring, Redundancy, Authentication |
| | Sybil | Probing, Authentication |
| | Worm holes | Authentication, Packet leashes by using geographic and temporal information |
| | Hello flood | Verify the bidirectional link, Authentication |
| | Acknowledgment spoofing | Authentication |
| Transport layer | Flooding | Client puzzles |
| | Desynchronization | Authentication |
| Application layer | Attacks on reliability and Clone attack: Clock skewing, Selective message forwarding, Data aggregation distortion | Unique pair wise keys and cryptographic approach. Authentication can be used to protect any data integrity Encryption is an effective approach for data confidentiality protection |

Table 1. Layer based attacks and possible security mechanisms in wireless sensor network

### 4.6 Node Replication Attack

It is quite simple attack. An attacker copies the node ID of an existing sensor node and adds to an existing sensor network. With node replication approach a sensor network's performance can severely disrupt. Packets can be corrupted or even misrouted. As a result this can form a disconnect network, false sensor reading etc. An attacker can copy cryptographic keys to the replicated sensors by physical accessing to the entire network. The attacker could easily manipulate a specific segment of the network by inserting the replicated nodes at specific network points, perhaps by disconnecting it altogether.

### 4.7 Attacks Against Privacy

Sensor network technology provides automatic data collection capabilities through efficient deployment of tiny sensor devices. These technology exhibits significant potential for abuse. As sensor network provides increased data collection capabilities, so privacy problems arise. The

41

main privacy problem is not that sensor network enable the collection of information. Direct site surveillance uses to collect much information from sensor networks. Sensor networks suffer privacy problems because they make large volume of information easily available through remote access. Adversaries can gather information at low-risk in anonymous manner because they need not be physically present to maintain surveillance.

### 4.8 Physical Attacks

Sensors networks typically operate in hostile outdoor environments. The sensor networks are highly susceptible to physical attacks, i.e. threats due to physical node destructions as sensors are small in size, deployed with the unattended environment. Physical attack destroys sensors permanently, so there are looses of cryptographic secrets, tamper with the associated circuitry, modify or replace sensors with malicious sensors under control of the attacker.

### 4.9 Wormhole Attack

Wormhole attack is a significant attack in WSN. This attack occurs at the initial phase when the sensors start to discover the neighbouring information [10].In this attack there is no need compromising a sensor in the network. In wormhole attack, the attacker records the packets (or bits) at one location in the network and tunnels those to another location. The tunnelling or retransmitting of bits could be done selectively. The simplest case of this attack is to have a malicious node forwarding data between two legitimate nodes. Wormholes convince distance nodes that they are neighbours, leading to quick exhaustion of their energy resources. An adversary situated close to a base station may be able to disrupt routing by creating a well-placed wormhole.

### 4.10 Spoofed, Altered or Replayed Routing Information

This is the most common direct attack against a routing protocol. This attack mainly targets the routing information exchanged between the nodes. In order to disrupt traffic in the network, an attacker may spoof, alter or reply routing information. These disruptions include the creation of routing loops, extending and shortening source routes, attracting or repelling network traffic from select nodes, partitioning the network, generating fake error messages and increasing end-to-end latency. Authentication is the standard solution for these attack i.e routers will only accept routing information from valid routers.

### 4.11 Selective Forwarding Attack

Multi-hop mode of communication is commonly preferred in WSN data gathering protocols. An assumption made in multihop networks is that all nodes in the network will accurately forward received messages. Selective forwarding attack is a situation when certain nodes do not forward many of the messages they receive. In this attack, malicious nodes may refuse to forward certain messages and simply drop them, ensuring that they are not propagated any further. When a malicious node behaves like a black hole and refuses to forward every packet she sees is a simple form of selective forwarding attack. This attack can be detected if packet sequence numbers are checked properly and continuously in a conjunction free network. Data packet sequence number in packet header can reduce this type attack. Vulnerability of this attack is logical.We have done comparative analysis of different threats in wireless sensor networks in table 2.

| Attacks | Principle[11] | Disadvantage | Network architecture | Vulnerability[14] | Security class[15] |
|---|---|---|---|---|---|
| Denial of Service (DoS) attacks | All Purposes | Eliminates the network's capacity | Traditional wireless sensor network | Logical | Interruption, interception, modification, fabrication |
| Wormholes | Unfairness, to be authenticated, to be authorized | Difficult to check routing information | Traditional wireless sensor network | Logical | Fabrication, interception |
| Spoofed, altered or replayed routing information | Unfairness | Generates false error messages, increase end-to-end latency, extend or shorten source routes | Large, distributed and traditional wireless sensor network | Logical | Fabrication, modification |
| Link layer Jamming | Disrupt communication | Exhausting node's resources, confusion and packets collision | Traditional wireless sensor network | Logical | Modification |
| Collision | Unfairness | Energy exhaustion, discarding packets, interefences, cost effective | Large and traditional wireless sensor network | Logical | Modification |
| Resource Exhaustion | Unfairness | Compromise availability, Resource exhaustion | Traditional wireless sensor network | Logical | Modification |
| Unfairness | Unfairness | Node's hungry to channel access, Efficiency and utility of service decrease | Traditional wireless sensor network | Logical | Modification |

| Acknowledge spoofing | Unfairness | False information to the neighboring nodes, packet loss,selective forwarding attack | Traditional and distributed wireless sensor network | Logical | Fabrication, modification |
|---|---|---|---|---|---|
| Sybil | Unfairness, To be authenticated, To be authorized | Launch threat to geographic routing protocol, Expensive in sensor network | Traditional wireless sensor network | Logical | Modification, Fabrication |

Principle[11]: Passive eavesdrop,disrupt communication, unfairness, to be authorized, to be authenticated

Vulnerabilities[14]: Physical (hardware) ,logical
Security class[15]: Interruption, interception, modification, fabrication

Table 2. Comparative analysis of different threats in wireless sensor networks

### 4.12 Passive Information Gathering

With a powerful receiver and well designed antenna an intruder can easily pickoff the data stream. An attacker locates the nodes and destroy them by interception of the messages containing the physical location of sensor networks[11]. An adversary can observe the application specific content of messages including message Ids, timestamps and other fields besides the location of sensor nodes.

### 4.13 Node Capturing

A particular sensor might be captured. Node capturing may reveal it's information including disclosure of cryptographic keys and an adversary can get information from the captured node [12].Node capturing compromises the whole sensor network and prevents from any communication. It launches a variety of insider attacks, software vulnerabilities.

### 4.14 False or Malicious Node

Attacks against security in WSN are caused by the insertion of false information from compromised nodes within the network [12].An intruder might add a malicious node to the system that feeds false data or prevents the passage of true data. The most dangerous attack is insertion of malicious node. Injecting malicious nodes destroy the whole network.

International Journal of Wireless & Mobile Networks (IJWMN) Vol. 6, No. 2, April 2014### 4.15 Energy Drain Attack

WSN is battery powered and dynamically organized. It is difficult or impossible to replace/recharge sensor node batteries. Attackers may use compromised nodes to inject fabricated reports into the network or generate large amount of traffic in the network as limited amount of energy available in sensor node. Fabricated reports will cause false alarms that waste real world response efforts and drain the finite amount of energy in a battery powered network. The network architecture of this attack is traditional wireless sensor network. This attack is possible only if the intruder's node has enough energy to transmit packets at a constant rate. This attack destroys the sensor nodes in the network, degrades performance of the network and ultimately splits the network grid and by inserting a new sink node takes control of part of the sensor network. Fabricated reports should be dropped en-route as early as possible to minimize the damage caused by this attack.

### 4.16 Acknowledgement Spoofing

Mostly sensor network routing algorithms rely on implicit or explicit link layer acknowledgements. An adversary can spoof link layer acknowledgments for 'overhead' packets addressed to neighbouring nodes due to the inherent broadcast medium. Protocols that choose the next hop based on reliability issues is susceptible packets being lost when travel along such links. The goal includes convincing the sender that a weak link is strong or that a dead or disabled node is alive. Acknowledgment spoofing attacks can be prevented by proper authentication for communication and using good encryption techniques [13].

## 5. CONCLUSIONS

Most of the attacks in wireless sensor networks are caused by the insertion of false information. For defending the inclusion of false reports by compromised nodes is required detecting mechanism. Developing such detection mechanism is great research challenge. All of the previously mentioned security threats i.e the HELLO flood attack, wormhole attack, sinkhole attack, Sybil attack serves one common purpose that is to compromise the integrity of the network they attack. In the past focus has not been on the security of WSNs. Security has become a major issue for data confidentiality as the various threats are arising. In this paper we have tried to present most of the security threats in WSN with extensive study.

**Authors**

**Sahabul Alam** received B.E in Information Technology from Biju Patnaik University of Technology, Rourkella, Orissa, India in 2005 and M.Tech in Computer Science and Engineering from National Institute of Technology, Hamirpur, Himachal Pradesh,India in 2010. His research interest is in wireless sensor network. He had been selected Maulana Azad National Fellowship by University Grants Commission,New Delhi,India.

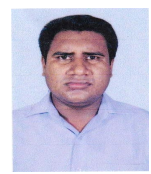

**Dr. Debashis De** is an associate professor in computer science and engineering department , West Bengal University of Technology, Kolkata, India. He received P.hd in nano technology engineering from Jadavpur University, India. He has published many research papers at International reputed journals and conferences. He is an author of different kinds of books. He received different kinds of awards from national and international level. His research interest is in mobile computing and nano technology.

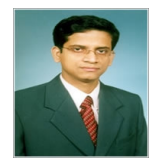